\documentclass[fleqn,12pt,twoside]{article}
\usepackage{espcrc1,epsfig,amssymb,amsbsy,xspace}

\newcommand{\AmS}{{\protect\the\textfont2
  A\kern-.1667em\lower.5ex\hbox{M}\kern-.125emS}}

     \newcommand{\pathnow}{}

\newcommand{\myfig}[7]{%
\begin{figure}[#7]
\vskip #5cm	\centerline{\hspace*{#1cm}
\epsfig{width=#2cm,angle=#4,figure=\pathnow #3.ps}
	}\vskip #6cm
                    }
\newcommand{\capt}[3]%
{\caption[{#2}]{\label{Fig:#1}#3}
}
\newcommand{\rf}[1]{Fig.\,\ref{Fig:#1}%
}
\def\qgp{quark--gluon plasma\xspace}
\def\QGP{\qgp}

\newcommand{\ress}[1]{subsection~\ref{ssec:#1}%
}

\newcommand{\req}[1]{Eq.\,(\ref{eq:#1})%
}

\newcommand{\beql}[1]{
	\begin{equation} \label{eq:#1}}

\newcommand{\beqarl}[1]{
	\begin{eqnarray} \label{eq:#1} }
\newcommand{\eeql}[1]{\label{eq:#1} \end{equation} 
} 
\newcommand{\eeqarl}[1]{\label{eq:#1} \end{eqnarray} 
}
\newcommand{\rt}[1]{table~\ref{Tab:#1}%
}
\def\beq{\begin{equation}}
\def\eeq{\end{equation}}

\def\beqar{\begin{eqnarray}}
\def\eeqar{\end{eqnarray}}
\def\bcite{\cite}
\def\agev{{$A$~GeV}\xspace}

\newcommand{\captt}[3]%
{\caption[{#2}]{\label{Tab:#1}#3%
}%
}
\newcommand{\lsec}[1]{\label{sec:#1} }

\newcommand{\lssec}[1]{\label{ssec:#1} 
 }

\def\etc{{etc.\xspace}}
\def\ie{{i.e.\xspace}}
\def\eg{{e.g.\xspace}}

\newcommand{\myfigd}[9]{%
\begin{figure}[#9]
\vskip #5cm	\centerline{\hspace*{#1cm}
\psfig{width=#2cm,angle=#4,figure=\pathnow #3.ps}
\hspace*{#7cm}
\psfig{width=#2cm,angle=#4,figure=\pathnow #8.ps}
	}\vskip #6cm
                    }
\title{Strangeness and Statistical QCD
}

\author{Johann Rafelski\address{Department of Physics,
University of Arizona, Tucson, AZ 85721\\
and\\ CERN-Theory Division, 1211 Geneva 23, Switzerland}\thanks{Supported by
grant DE-FG03-95ER40937 from the U.S. Department of
Energy.}
and
Jean Letessier\address{
Laboratoire de Physique Th\'eorique et Hautes Energies\\
Universit\'e Paris 7, 2 place Jussieu, F--75251 Cedex 05.
}\thanks{LPTHE, Univ.\,Paris 6 et 7 is:
Unit\'e mixte de Recherche du CNRS, UMR7589.}
}
\begin{document}

\maketitle

\begin{abstract}
We discuss properties of statistical QCD relevant in 
Fermi phase space model analysis of strange
hadron production experimental data. We argue that the 
analysis results interpreted using established statistical 
QCD properties are demonstrating formation of the color
deconfined state of matter in relativistic 
heavy ion collisions at highest CERN-SPS energies 
and at BNL-RHIC,  comprising
deconfined matter composed of nearly massless quarks and gluons, in 
statistical equilibrium.
\end{abstract}

\vskip -13cm \ \hfill CERN-TH/2001-347 \vskip 13cm
\section{INTRODUCTION}
More than 20 years ago, several theoretical
meetings held at the University Bielefeld have laid the 
foundation for the rapid analytical (perturbative QCD) and
numerical (lattice QCD) development of Statistical 
Quantum Chromodynamics. We will illustrate in this paper, using
as an example the strangeness signature of quark-gluon 
(color) plasma (QGP),
the relevance of this work in the analysis of experimental
data, and the resulting identification of quark-gluon plasma
formation in relativistic heavy ion collisions.

The discovery of QGP, unlike the discovery of a new
particle, requires a change in the understanding of
fundamental hadronic degrees of freedom. The
signature of QGP is very hard to identify, since the
final state observed in the laboratory always 
consists of the same particles, irrespective of the
transitional presence of the deconfined state. What changes is 
 the detailed composition of the observed 
produced particle abundance.
We consider here the disappearance of 
strangeness suppression, accompanied by a changed pattern
of (enhanced) strange antibaryon production, with 
reestablished baryon-antibaryon symmetry of thermal
spectra. All these phenomena offer compelling evidence for 
quark-gluon plasma 
formation \cite{Raf81,Raf81b,Raf81f,Raf82c,Raf82b,Koc86,Koc86d,Raf82a,RD83}.

To demonstrate the formation of QGP, we
 determine the physical properties of the
source of hadronic particles we are observing, which
we associate with the properties of statistical
QCD. We show that the anomalous yields of strange
hadrons originate from a state of matter with 
physical properties of a quark-gluon fireball. 
We begin with a short survey of statistical QCD results
we employ, and follow this with a summary of results on 
hadron production and their interpretation in terms of the 
QGP state.

\section{STATISTICAL QCD}
\label{sqm}
\subsection{Running coupling constant of QCD}
\label{alpharun}
For the purpose of quark--gluon plasma studies,
we require the strength  of QCD interaction, and the 
magnitude of quark mass as function of energy scale. 
The simplest way to obtain these results is to 
integrate the first order renormalization group
equations using the perturbative definition  of
the $\beta,\gamma$ functions,
 given an initial value of $\alpha_s(M)$ and $m_i(M)$.
For the determination of the coupling constant, it has become 
common to refer to the value of \index{$\alpha_{s}(M_Z)$}
$\alpha_{s}(M_Z=91.19\,\mathrm{GeV})$. We use as the central
value  $\alpha_{s}(M_Z)=0.1182\pm0.002$.

To obtain the solid line in figure \rf{alfaIN} (left) the 
four-loop $\beta$-function was used,
in the $\overline {MS}$ modified minimum subtraction scheme.  
The dotted lines  demonstrate considerable sensitivity 
to the initial value  $\alpha_{\rm s}(M_Z)$. 
If  $\alpha_{\rm s}(M_Z)$ were larger, the evaluation of the 
coupling strength in the `low' energy domain $\mu\lesssim 1$ GeV, of 
interest in QGP studies, would be divergent,
see dotted lines in \rf{alfaIN}  (left)  above 
the solid line. The criticism of the perturbative study of strangeness 
production derives from the belief that the strength of the QCD interaction
is larger than is now known: to put it differently,  an essential prerequirement 
for the perturbative theory of strangeness production in QGP, is the 
relatively small experimental value $\alpha_{\rm s}(M_Z)\simeq 0.118$, 
which has been experimentally established in recent years. It is interesting 
to note in figure \rf{alfaIN} (left) 
that a 20\% reduction in $\alpha_{\rm s}(M_Z)$  leads 
to a `good' $\alpha_{\rm s}(0.1\,\mbox{GeV})<1$. 

\myfigd{1.5}{10.5}{ALSMU4BB}{0}{-3.}{-3.}{-3.}{PTHERMalphaB}{tb}
\capt{alfaIN}{$\alpha_{\rm s}(\mu)$ for a variety of initial conditions}{
\small
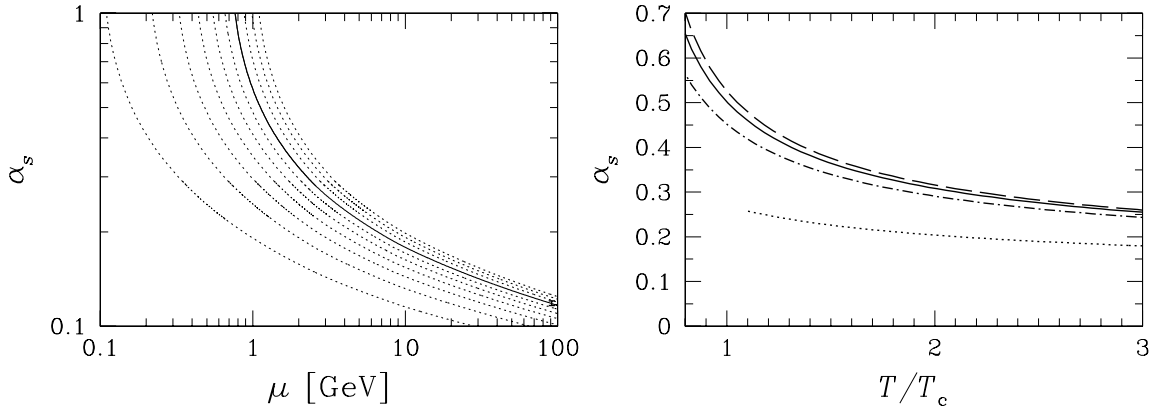
Left: $\alpha^{(4)}_{\rm s}(\mu)$ as function of energy scale $\mu$ 
for a variety of initial conditions. 
Solid line: $\alpha_{\rm s}(M_{Z})=0.1182$ (see the experimental point,
which includes the error bar at $\mu=M_Z$;
dotted lines: sensitivity  to variation of the initial condition.\\
Right: $\alpha_s(2\pi T)$ for $T_{\rm c}=0.160$\,GeV. 
Dashed line: $\alpha_s(M_Z)$ = 0.119; solid line = 0.1181;
dot-dashed line = 0.1156.
Dotted line: approximate 2-loop solution, given in Eq.\,\ref{eq:Lambdarun2},
with choice $\Lambda=150$ MeV.
}
\end{figure}

When studying thermal processes in QGP at temperature $T$, the proposed 
interaction scale is $\mu\equiv 2\pi T\simeq 1\,\mbox{GeV}\ T/T_{\rm c}$\,,
for $T_{\rm c}\simeq 160$\,MeV.  In \rf{alfaIN} (right), the solid line 
corresponds to an exactly computed  $\alpha_s$ with physical quark 
thresholds, evaluated at the thermal scale, and expressed in terms of 
$T/T_{\rm c}$. The range of experimental uncertainty in $\alpha_s(T)$, 
due to uncertainty in $\alpha_s(M_Z)$, is
delimited by  dashed and dot-dashed lines bordering the solid
line. A popular approximation of $\alpha_s$,
\beql{Lambdarun2}
\overline{\alpha}_s^{(2)}(\mu)\simeq{2 \over b_0 \bar{L}}\left[1-
\frac{2 b_1}{b_0^2}\frac{\ln\bar{L}}{\bar{L}}\right]\,,\quad
\bar{L}\equiv\ln(\mu^2/\Lambda^2)\,,
\eeq
agrees, using the standard value 
$\Lambda^{(5)}=205\pm25$ MeV with exact solution for  $\mu>2m_b$. 
Presence of lighter quark loops leads to a major deviation at
scales of interest to us. \req{Lambdarun2} 
is represented in \rf{alfaIN} (right) by the dotted line, it
misses the exact result by factor 2 for 
$T_{\rm c}<T<1.75T_{\rm c}$, the 
effective range of observables emerging in SPS and
RHIC experiments. The experimental error in determination of $\alpha_s$ 
is today considerably smaller.

The high sensitivity of physical observables to $\alpha_s$, 
which depend on $\alpha_s^n, n\ge 2$, makes it 
imperative that we do not rely on this  approximation (dotted line in 
\rf{alfaIN} (right)). Yet a
fixed value  $\alpha_s=0.25$ (instead of $\alpha_s= 0.5$) 
derived from this approximation 
is still often used  in a study of  the QGP phase
properties, energy loss of parton jets, charmed quark thermalization,
strange quark thermal production, \etc\ Such treatment of QCD interaction 
underestimates by as much as factor four the interaction with the QGP 
phase, and thus the speed of these processes. In most cases,
this mundane factor matters, and we see that 
accurate evaluation of $\alpha_s$ at the
appropriate physical scale is required.

\subsection{Quark--gluon plasma pressure and phase boundary}
The partition function (\ie, pressure) of the quark--gluon phase  is obtained 
after we combine quarks, gluons and vacuum contributions:
\beqarl{ZQGPL1}
\frac{T}{V}\ln{\cal Z}_{\mathrm{QGP}}
&\hspace{-0.3cm}\equiv &\hspace{-0.3cm}P_{\mathrm{QGP}}=
-{\cal B}\!+\!\frac{8}{45\pi^2}c_1(\pi T)^4  
\!+\!\frac{n_{\mathrm f}}{15\pi^2}
\!\left[\frac{7}{4}c_2(\pi T)^4\!+\!\frac{15}{2}c_3\left(
\mu_{\mathrm q}^2(\pi T)^2 \!+\! \frac{1}{2}\mu_{\mathrm q}^4
\right)\right].
\eeqar
We have inserted here the appropriate quark ($n_{\rm f}=2$) 
and gluon degeneracy.
The interactions between
quarks and gluons manifest their presence aside of the 
vacuum structure effect ${\cal B}$,
in the three coefficients $c_i\ne 1$, see  \cite{Chi78}:
\beql{ICZQGP1}
c_1= 1-\frac{15\alpha_s}{4\pi}+ \cdots\,,\quad
c_2= 1-\frac{50\alpha_s}{21\pi}+ \cdots\,,\quad
c_3=1-\frac{2\alpha_s}{\pi}+ \cdots\,.
\eeq

While higher order terms in
the perturbative expansion have been obtained,
they suggest lack of convergence of the expansion
scheme formulated today. Given that the lowest order 
terms are consistent with thermal lattice results
we do not pursue this issue further. Rather, 
we show for which parameter set the lattice and 
perturbative statistical QCD results agree with
each other and we use these results to understand 
the properties of a QGP fireball.
To reproduce the lattice results, near to $T=T_{\rm c}$,
the only choice we can make is in value of the bag constant, 
${\cal B}=0.19\,{\mbox{GeV}}/{\mbox{fm}}^3$\,.
\myfig{-0.2}{12}{PLMUPLIQ3B}{0}{-2.}{-1.6}{tb}
\capt{PLMUPLIQ}{P=0 in mu-T and at finite v}{
\small
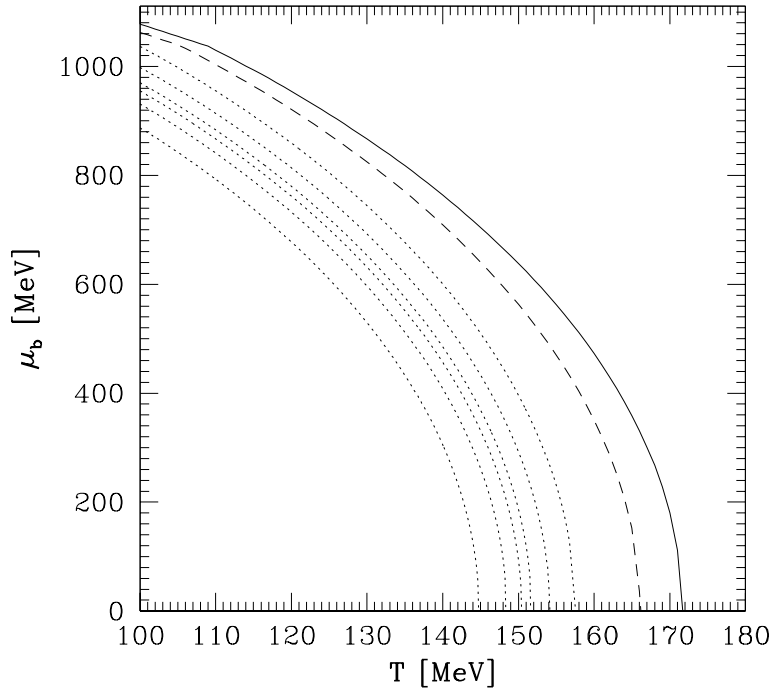
Pressure balance in 
$\mu_{\mathrm b}$--$T_{\rm c}$-plane. Dotted(from
right to left):  $P=0$ at expansion velocity 
$v^2=0, 1/10, 1/6, 1/5, 1/4$ and $1/3$. Solid line, the
phase boundary between hadron gas and QGP, with point hadrons, dashed 
the phase boundary with finite size hadrons. 
}
\end{figure}

Dotted lines in  \rf{PLMUPLIQ} show the QGP phase pressure  condition
$P_{\mathrm{QGP}}\to 0$ for different velocities of expansion, employing 
an extrapolation to finite baryochemical potential \cite{Ham00}. The
solid line denotes, in the 
$\mu_{\mathrm b}$--$T$ plane the pressure balance 
between an hadronic point particle gas
and a stationary  quark--gluon phase.
Dashed line shows the phase boundary with an hadronic gas made of finite size hadrons.
However, the fluid flow motion of quarks and gluons 
expands the domain of deconfinement exercising against the vacuum 
force originating in the collective 
expansion velocity $\vec v_{\rm c}$ \cite{Raf01}.  

The condition $P=P_{\mbox{\scriptsize p}}-{\cal B}=0$, including the effect of 
motion  reads \cite{Raf00},
\beql{BPeqvS}
{\cal B}=P_{\mbox{\scriptsize p}}+
      (P_{\mbox{\scriptsize p}}+\epsilon_{\mbox{\scriptsize p}})
\frac{\kappa  v_{\mbox{\scriptsize c}}^2}{1-v_{\mbox{\scriptsize c}}^{2}}\,,\qquad 
\kappa=\frac{(\vec v_{\mbox{\scriptsize c}}\cdot \vec n)^2}{v_{\mbox{\scriptsize c}}^2}\,,
\eeq
where subscript p refers to particle pressure and energy, and
where we introduced the geometric factor $\kappa$ which
characterizes the angular relation between the surface normal vector and flow direction. 

Expansion beyond  ${P}\to 0$ is in general not possible. 
A fireball surface region  which reaches condition \req{BPeqvS}
and continues to flow outwards
must   be torn apart. This is a collective instability and the 
ensuing disintegration of the fireball matter should be  very rapid. 
Thus we find, that a rapidly evolving fireball which supercools into the
domain of negative pressure is in general highly unstable, and we expect that
a sudden transformation (hadronization)
into confined matter can ensue in such a condition. It is important to note 
that the situation we described could only arise  since the vacuum pressure 
term is not subject to flow and always keeps the same value.
Looking at the high flow velocity curves in  \rf{PLMUPLIQ} we see that 
an exploding QGP  fireball only contained by the vacuum can supercool to $T\simeq 145$ MeV.

\subsection{Thermal mass and QGP properties}
It has been shown that it is  also  possible to reproduce
the  lattice results using fine tuned  thermal masses 
(see table I in \bcite{Pes00}). 
In \rf{PLMGMQ},  we show the light quark (solid thick line)
and gluon (dashed thick line) thermal masses which were
fitted to the lattice QGP data. The perturbative  thermal masses,
definitions slightly modified compared to \bcite{Wel82,Wel82a}, are for quarks and gluons,
\beql{mqthermal}
(m_{\rm q}^{\rm T})^2=
               \frac{4\pi}{3}\alpha_s T^2\,,\qquad 
(m_{\rm g}^{\rm T})^{\,2}=
           2\pi\alpha_s T^2 \left(1+\frac{n_{\rm f}}{6}\right)\,,
\eeq
and are also shown in \rf{PLMGMQ}, thin lines (dashed for gluons) obtained
using  $\alpha_s(\mu=2\pi T)$, see \rf{alfaIN} (right).

\myfig{0}{12}{PLMGMQB}{0}{-4.6}{-1.6}{tb}
\capt{PLMGMQ}{thermal mass}{
\small
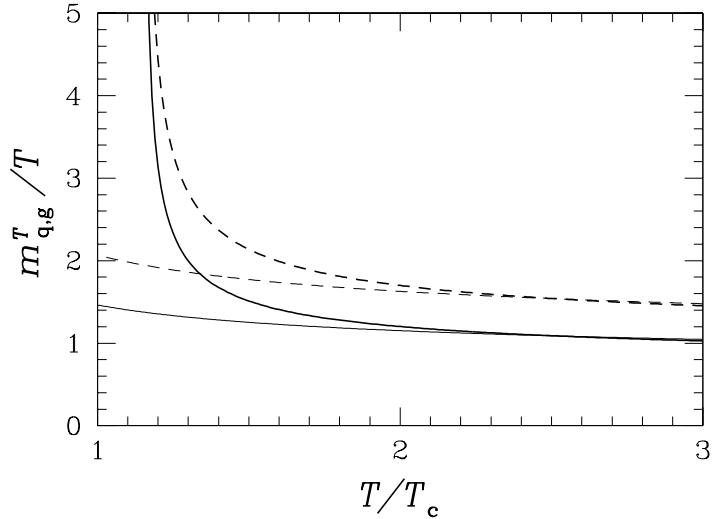
Thermal masses fitted to reproduce Lattice-QCD 
results \protect\cite{Pes00},
thick solid line for quarks, and thick dashed line for gluons. 
Thin  lines, perturbative QCD masses for $\alpha_s(\mu=2\pi T)$.
\index{thermal mass}}
\end{figure}

We conclude that the thermal masses required to describe the 
reduction of the number of degrees of freedom for $T>2 T_{\rm c}$ are just 
the perturbative QCD result. Importantly, 
this means that thermal masses  express,
in a  different way, the effect of  perturbative QCD, and thus for
$T>2T_{\rm c}$, we have the option to use \req{ZQGPL1}, or the more
complex thermal mass approach.

The discussion of the properties of QGP is a wide subject which is 
beyond the scope of this report, and we conclude showing the 
entropy density as function of temperature in \rf{Lentro}.
The solid line is for the case of equilibrated quark-light
glue system, in the limit of vanishing
chemical potential.  We note that initially
the entropy rises faster than the asymptotic $T^3$ 
behavior, since the QCD interactions weaken, and
there is an increase in the effective number
of acting quark and gluon degrees of freedom. 
Thus, the drop in entropy density is considerable
when plasma cools and approaches  hadronization condition. In order to 
preserve the entropy content in the fireball 
when the system expands, from $T\simeq 300$ MeV 
towards  $150$ MeV, a 
volume dilution by a factor nine must occur.
The `pure glue' case (short-dashed line) contains as expected
 about half of the entropy when
comparing at equal temperature, the addition
of strangeness  adds at the level of 10\% to the entropy content,
at a given temperature. However, as strangeness is produced temperature
of the isolated fireball cools, and thus there entropy production 
is even smaller.

\myfig{0}{12}{PLLENTROB}{0}{-4.6}{-1.7}{tb}
\capt{Lentro}{Entropy density in  QGP}{
\small
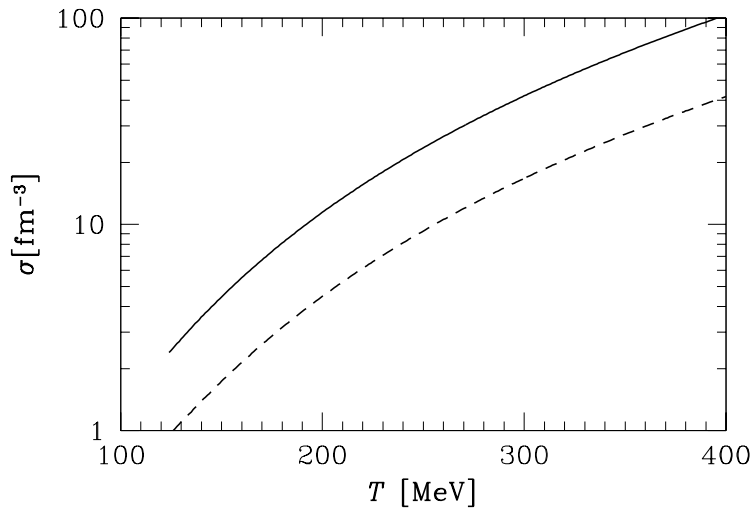
Entropy density in chemically equilibrated  QGP
at $\lambda_{\rm q}=1$  as function of temperature. 
Solid line $n_{\rm f}=2$, long dashed line  $n_{\rm f}=2.5$,
short dashed line `pure glue'  $n_{\rm f}=0$.
\index{entropy density!quark--gluon liquid}}
\end{figure}

\section{STRANGENESS PRODUCTION IN COLLISION PROCESSES}
\subsection{Remarks about thermal strangeness production}
 \lssec{thermalpart}  
Strangeness, and more generally  heavy flavor quarks, can be
produced either in the first interactions of colliding matter,
or in the many ensuing less energetic collisions.
Strange quark mass $m_{\rm s}$ is comparable in 
magnitude to the typical temperatures reached in heavy ion interactions,
 and the numerous `soft' secondary parton collisions 
dominate the production of strangeness, and naturally, of the light $\rm u,d$
flavors.  

At the time strange flavor approaches chemical equilibrium in soft collisions,
the back reaction is also relevant. The quantum mechanical matrix element 
driving a two-body reaction must be, channel by channel, the same for forward and
backward going reactions
due to time reversal symmetry.  The actual rate of reaction differs,
since there are usually considerable differences in statistical and phase space 
factors. However, the forward an backward reactions will
balance when equilibrium particle yields are established.
This principle of detailed balance can sometimes  be used to 
evaluate one of the two reactions.

This is not the place to repeat the well known details of the kinetic
theory of strangeness production, \cite{Raf82a} based on the perturbative
cross sections \bcite{Com79}, evaluated with running QCD parameters,
\cite{Let96}. However, we note that the 
use of scale dependent QCD parameters, $\alpha_s$ and $m_{\rm s}$,
with $\mu\propto \sqrt{s}$
amounts to a re-summation of many QCD diagrams comprising vertex, and 
self energy corrections. A remaining shortcoming of thermal production
evaluation  is that  up to day, there has not  been a study of the 
next to leading order final state accompanying gluon
emission in thermal processes, \eg, $\rm gg\to s\bar s+g$. 
In direct parton induced reactions, this next to leading order 
effect enhances the production rate by a factor {$K=1.5$--$3$}. This 
may cause a corresponding increase also in the thermal rate of production, and
thus a reduction in the thermally computed 
chemical equilibration time of strangeness and charm.
The  magnitudes (up to 0.4\,mb) of both types of 
reactions considered, quark fusion and gluon fusion,  are similar.
Gluons dominate flavor production, because they are so much more
likely to collide in the appropriate production channel. 

Given the cross section and collision frequency, the  relaxation 
constant $\tau_{\rm s}$ of chemical equilibration can be evaluated and 
the result is that it takes less than 2 fm/c to equilibrate strangeness
if $T>300$ MeV, and somewhat less time, in the event that 
next to leading order effects play an important role. Despite this
clear indication of near strangeness equilibration in SPS and
RHIC environment,
there is considerable difference in the computational  
results \bcite{Bir93,Sri97,Sri97b,Raf99a},
originating in differing assumptions about degree of chemical equilibration 
of gluons in the early stage of the QGP formation. The relaxation time 
is 4 times larger should gluons be only 50\% equilibrated.

\subsection{Strangeness background} 
     \lssec{Wrob}
One could  argue that 
if strangeness is to be used as a diagnostic tool 
of \qgp, we need to understand this background production rate 
of strange hadrons. In that context, we are interested to measure how 
often, compared to light quark pairs, strange quarks are made.

Wr\'oblewski proposed  to consider the  strangeness suppression 
factor  \bcite{Wro85}:
\beql{wro1}
W_{\rm s} = 
{ {2 \langle\rm  s   \bar s \rangle} \over 
{\langle\rm   u \bar u \rangle + \langle\rm  d\bar d \rangle} }\,.
\eeq
Only newly made $\rm s\bar s$, 
$\rm u\bar u$ and $\rm d\bar d$ quark pairs are counted. 
If strangeness were to be as easily produced as light $\rm u,d$ quarks, we 
would find $W_{\rm s}\to 1$. To obtain the experimental value for $W_{\rm s}$, 
a careful study of produced hadron yields is required. We cite below results
obtained using a semi-theoretical method \bcite{Bec98}, in
which numerous particle yields are described within the framework of 
a statistical model. In elementary collisions 
$\rm pp$, $\rm p\bar p$, $\rm e^+e^-$, a value  $W_{\rm s}\simeq 0.22$ is obtained,
strangeness is thus relatively strongly suppressed. On the other hand, 
in nuclear A--A' collisions $W_{\rm s}$ more than  doubles
compared to p--p interactions, considered at the same energy. 

To explain the 2--2.5-fold strangeness yield increase in a kinetic model of 
particle production requires
a shift towards strangeness production of all particle formation processes. 
In other words, when modeling the enhanced strangeness yields within a variety of
approaches,  in each model a new reaction mechanism must
be introduced that favors strangeness over none-strange processes.
Even at this relatively elementary level of counting  hadron
abundances, new physics must be introduced. In a model with 
the deconfined phase this new reaction mechanism is due to the 
presence of mobile gluons, which are most
effective in making strange quark pairs. Moreover, as the conditions
created in the QGP become more extreme with collision
energy, \eg, initial temperature exceeding
substantially the strange quark mass, we expect an increase in 
$W_{\rm s}$.

Generally hadron cascade models tuned to
produce enough singly strange hadrons, predict wrong abundances of the rarely produced 
particles such as $\overline\Xi$, $\overline\Omega$.
We are not aware of any kinetic hadron  model with or without
`new physics' that is capable to reproduce the pattern of rare
hadron production, along with  enhancement of strangeness and hadron 
multiplicity. Moreover, if rapidity
spectra are modeled,  usually the transverse momentum
spectra are incorrect, or vice-versa. It thus seems impossible, in a 
collision model based on confined hadron interactions, to find
sufficiently many hadron--hadron collisions to occupy by hadrons
the large phase space (high $p_\bot$, high $y$) filled by 
products of nuclear collision. If indeed a non-QGP
reaction picture exists to explain heavy ion collision data,
the current situation suggests that some essential reaction
mechanism has been overlooked for 20 years. In short, hadron
models need a lot more effort to reach satisfactory agreement with the
experimental results, even regarding rather simple observable such as
hadron multiplicities, transverse energy production, strangeness 
yield. Strangeness 
production in thermal processes in hadronic gas 
were studied carefully  \bcite{Koc85,Koc86},
and  the time scales involved  established. 
It is impossible to explain the observed pattern
of strange hadron production in kinetic thermal model based on
confined hadron interactions. 

Some researchers abandon the kinetic, \ie, collisional theory of 
particle production, and focus solely on the experimental fact that
the observed hadronic multiplicities are result of a pre-established 
statistical distribution near to thermal and chemical 
equilibrium, which works
quite well. However, it was already pointed out 15 years 
ago that such a result  can  be naturally explained in terms
of a dynamic theory of a transient deconfined state 
hadronizing in a coalescence model \bcite{Koc86d}. A detailed 
study of the subtle deviations in hadron yields 
 from precise statistical equilibrium yields 
allows to understand the hadronization mechanism \bcite{Bir00}, and 
therefore ultimately also to explore the properties of
the hadronizing QGP state. 

\section{HADRON FREEZE-OUT}
\lsec{fermisec}
\subsection{Chemical nonequilibrium in hadronization}
\lssec{SudHad}
An extra reaction step 
`hadronization' is required to connect the properties 
of the deconfined quark--gluon matter  fireball,
and the experimental apparatus.
In this process, the quark and gluon content of the 
fireball is transferred into ultimately free
flowing hadronic particles. In hadronization,  
gluons fragment into quarks, and quarks coalesce into
hadrons. Color `freezes', 
\qgp\ excess entropy has to find a way to get away, so any additional
production is hindered. It is far from obvious that  hadron 
phase space (so called `hadronic gas')
be used consistently to describe the physics of thermal hadronization, and
we establish now the consistency criterion assuming that 
entropy production is small, or even null, in 
hadronization of entropy rich thermal QGP into entropy poor hadron
phases space.  

We 
consider the Gibbs--Duham relation for a unit volume, 
$\epsilon+P=T\sigma+\mu \nu$\,,
and combine it with the instability condition of dynamical expansion, 
 \req{BPeqvS}:
\beql{BPeqvS1} 
0=P|_{\rm h}+(P|_{\rm h}+\epsilon|_{\rm h}) {\kappa v_{\rm c}^2\over 1-v_{\rm c}^2}\,,
\quad
\left.{\epsilon\over \sigma}\right|_{\rm h}=
\left(T|_{\rm h}+\left.{\mu_{\rm b}\nu_{\rm b}\over \sigma }\right|_{\rm h}\right)
\left(1+{\kappa v_{\rm c}^2\over 1-v_{\rm c}^2}\right)\,.
\eeq
Using global variables we obtain \bcite{Raf00}:
\beql{EBSfinal}
\left.{E\over S}\right|_{\rm h}=
(T|_{\rm h}+\delta T|_{\rm h})\left(1+{\kappa v_{\rm c}^2\over 1-v_{\rm c}^2}\right)\,,\quad
\delta T=\mu_{\rm b}\frac{\nu_{\rm b}}{\sigma}
=\frac{\mu_{\rm b}}{S/b}\,.
\eeq
For RHIC,  we have 
$ \delta T|_{\rm h} < 0.4$ MeV,
considering that $\mu_{\rm b}<40$ MeV and $S/b>100$; at top SPS  energy,
we have $\mu_{\rm b}\simeq 200$--250 MeV and $S/b\simeq 25$--45, and thus, 
 $\delta T|_{\rm h} \simeq 5$ MeV. 

The particular usefulness of \req{EBSfinal} comes from the observation
that it implies:
\beql{EBSineq}
\left.{E\over S}\right|_{\rm h} > T|_{\rm h}\,.
\eeq
This is a near equality since
the geometric emissivity factor $\kappa$ is
positive and small, especially so at RHIC, and  $v_{\rm c}^2<1/3$.

The  Gibbs--Duham relation implies:
${E/ S}+{PV/ S}=T+\delta T>T$\,.
One would think that the $PV$ term is small, since
the pressure is small as the lattice calculations suggest,
$\epsilon/P\to 7$. However, this effect
can be compensated by large volume of hadronization. The volume 
is a directly measured quantity. The HBT results
place  a very severe  constraint on the emitter size  of the pion 
source. Another 
way, to constraint the magnitude of $V$, is to relate it to the 
total yield of pions. With these constraints, the $PV$  term is negligible.
In  \req{EBSfinal}, due to super cooling, $P$ was  negative, and contributed
 the $\kappa$ term on the right hand side.  

Neglecting the influence of $PV$,  we obtain just as in \req{EBSineq}:
${E/S} \simeq T$\,.
In \req{EBSineq}, we have a universal hadronization constraint,
which should be satisfied in models claiming to describe hadron production
in relativistic heavy ion collisions. 
This constraint is, in our
experience, difficult if not impossible to satisfy in chemical equilibrium
statistical  hadronization models. The reason is  that, in 
such an approach, a rather high 
temperature $T\simeq 175$ MeV is required to accommodate the high
intrinsic entropy content of 
hadron source, but this does not drive up sufficiently the 
energy content, since for pions in equilibrium $E/S<T$. 
In the super saturated ($\gamma_{\rm q}\sim 1.6$) pion gas $E/S>T$.

\subsection{Phase space and parameters}
\lssec{Phasespace}
Our approach  is in its
spirit  a generalization of Fermi's statistical model of 
hadron production \bcite{Fer50,Fer53}, in that the yield
of hadrons is solely dictated by the study of the 
magnitude of the phase space available.

The relative number of  final state hadronic particles
freezing out from, {\eg}, a thermal quark--gluon source, is obtained
noting that the  fugacity $f_i$ of the $i$-th emitted  composite  
hadronic particle containing $k$-components is derived 
from fugacities $\lambda_k$ and phase space occupancies $\gamma_k$:
\beql{abund}
N_i\propto e^{-E_i/T_{\rm f}}f_i=e^{-E_i/T_{\rm f}}\prod_{k\in i}\gamma_k\lambda_k.
\end{equation}
In most cases, we study chemical properties of 
light quarks $\rm u,d$ jointly, though on occasion, we 
will introduce the isospin asymmetry.
As seen in \req{abund}, we study particle 
production in terms of five statistical parameters
$T, \lambda_{\rm q}, \lambda_{\rm s}, \gamma_{\rm q},  \gamma_{\rm s}$. In addition,
to describe the shape of spectra, one needs  matter flow velocity
parameters, these become irrelevant when only total particle abundances
are studied, obtained integrating  all of phase space, or equivalently
in presence of strong longitudinal flow, when we are looking at a
yield per unit of rapidity. 

Assuming a QGP source, several of the statistical parameters
have natural values:
\begin{enumerate}
\item $\lambda_{\rm s}$: The value of strange quark 
fugacity $\lambda_{\rm s}$ can be obtained  
from the  requirement that strangeness  balances,
$\langle n_{\rm s}-n_{\bar s}\rangle=0\,,$
which for a source in which all $\rm s,\bar s$ quarks are unbound and 
have symmetric phase space, implies $\lambda_{\rm s}=1$\,.
 However, the Coulomb distortion
of the strange quark phase space plays an important role in the
understanding of this constraint for Pb--Pb collisions, 
leading to the Coulomb-deformed value 
$\lambda_{\rm s}\simeq 1.1$\,.
\item $\gamma_{\rm s}$: The strange quark phase space occupancy 
$\gamma_{\rm s}$ can be computed within the framework 
of kinetic theory and  $\gamma_{\rm s}\simeq 1$\,. Recall that the 
the difference between the two different types of chemical parameters
$\lambda_i$ and $\gamma_i$ is that the phase space
occupancy  factor $\gamma_{i}$ regulates the number of pairs of flavor `$i$', 
and hence applies in the same manner to particles and antiparticles, while
fugacity $\lambda_i$ applies only to particles, while $\lambda_i^{-1}$ is
the antiparticle fugacity. 
\item $\lambda_{\rm q}$: The light quark fugacity $\lambda_{\rm q}$, 
or equivalently, the baryochemical potential $\mu_{\rm b}$,
regulate the baryon density of the fireball and hadron  freeze out. 
This density can vary 
dependent on the energy and size of colliding nuclei, and   
the value of $\lambda_{\rm q}$ is not easily predicted. 
However, it turns out that this is the most precisely 
measurable parameter, with everybody obtaining the same 
model independent answer, for it directly enters all very abundant
hadrons. Since $T$ differs depending on strategy of the analysis, the
value of $\mu_{\rm b}$ is not so well determined and we recommend that
$\lambda_{\rm q}$ be cited instead of  $\mu_{\rm b}=3T\ln\lambda_{\rm q}$.
\item $\gamma_{\rm q}$: The equilibrium phase space occupancy of light quarks 
$\gamma_{\rm q}$ is expected to  significantly exceed unity 
to accommodate the excess  entropy content in the 
plasma phase, 
$\gamma_{\rm q}\le \gamma_{\rm q}^{\rm c}=e^{m_\pi/2T_{\rm f}}\simeq 1.6$. 

\item $T_{\rm f}$: The freeze-out temperature $T_{\rm f}$ is  expected to be 
within 10\% of the Hagedorn temperature 
$T_H\simeq 160$\,MeV, which characterized particle production in 
proton-proton reactions. 
\item Turning now to the  flow parameters:
The  collective  expansion velocity $v_{\rm c}$ is expected 
to remain near to the relativistic sound velocity, 
$v_{\rm c}\le 1/\sqrt{3}$,
the natural flow speed of information in the QGP phase. There
is a longitudinal velocity which is needed to  describe
rapidity spectra, and there is a hadronization surface
motion, aside of many further parameters one may wish to use
to model profile of velocity of flowing matter.
\end{enumerate}

The resulting yields of final state hadronic particles are most 
conveniently characterized taking the Laplace transform of the 
accessible phase space. This approach generates a  function which,
in its mathematical properties, is identical to 
the partition function. For example,
for the open strangeness sector, we find (with no flow),
\beql{4abis}
{\cal L}\left[e^{-E_i/T_{\rm f}}\prod_{k\in i}\gamma_k\lambda_k\right] 
 \propto \ln{\cal Z}_{\rm s}^{\rm HG}\,.
\eeq

It is important to keep in mind that:
\begin{enumerate}
\item[a.] \req{4abis} does not require formation of a phase comprising a
gas of hadrons,  but is not inconsistent with such a step in evolution 
of the matter;  \req{4abis} describes  not a partition function, but just 
a look-alike object arising from the  Laplace transform of the accessible 
phase space; 
\item[b.] the final particle abundances measured in an experiment 
are obtained after all unstable hadronic resonances `$j$'
are allowed to disintegrate, contributing to the yields of
stable hadrons;
\item[c.] in some experimental data, it is important to distinguish 
the two light quark flavors, for example experiments are only sensitive
to $\Xi^-$ and not $\Xi^0$ and  an average over isospin does not
occur.
\end{enumerate}

 The unnormalized particle multiplicities arising are obtained 
differentiating \req{4abis} with respect to particle 
fugacity. The relative particle yields are simply 
given by ratios of corresponding chemical factors, weighted with
the size of the momentum phase space accepted by the experiment.
The ratios of strange antibaryons to strange baryons {\it
of same particle type\/} are, in our approach, simple functions of the quark 
fugacities.

\subsection{Strange hadrons at SPS}
\lssec{noneqchem}\lssec{partyield}\lssec{spsec}\lssec{Chemicalfreeze}
We expect in sudden hadronization 
chemical non-equilibrium at hadron freeze-out. 
For strangeness, $\gamma_{\rm s}\ne 1$, has been seen early on in
experimental data, \bcite{Raf91a}.
Full chemical non-equilibrium has been first noted in the study of the 
S--Au/W/Pb collisions at 200$A$ GeV \bcite{Let99}. Fitting 
the yields of hadrons observed, it has been reported that 
the statistical significance
increased when chemical non equilibrium was introduced. 
The statistical significance is derived from the statistical error: 
\beql{chi2}
\chi^2\equiv\frac{\sum_j({R_{\rm th}^j-R_{\rm exp}^j})^2}{
({{\Delta R _{\rm exp}^j}})^2}\,.
\end{equation}
It is common to normalize the total error $\chi^2$ by the 
difference between the number of data points and parameters used,
the so called `dof' (degrees of freedom) quantity.
For systems we study, with 
a few degrees of freedom (typically 5--15),  a  statistically 
significant fit requires that $\chi^2/$dof $<1$\,.
For just a few `dof', the error should 
be as small as $\chi^2/$dof$<0.5$.
The usual  requirement  $\chi^2\to 1$ is  only applying
for infinitely large `dof'.

Turning to the Pb--Pb system at 158$A$ GeV collision energy,
we consider  particle listed  in \rt{resultpb2}, top section from the
experiment WA97, for $p_\bot>0.7$\,GeV, within a narrow
$\Delta y=0.5$ central rapidity window. Further below 
are shown  results from the large  acceptance experiment NA49, 
extrapolated by the collaboration to full $4\pi$ phase space coverage. 
The total error $\chi^2$
for the  two result columns is shown at the bottom of this table
along with the number of data points `$N$', parameters `$p$' 
used, and number of (algebraic) redundancies `$r$' connecting the 
experimental results. For $r\ne 0$, it is more appropriate 
to quote the total  $\chi^2$, since the 
statistical relevance condition is more difficult to 
establish given the constraints, but since
 $\chi^2/(N-p-r)<0.5$, we are certain to have a valid description of
hadron multiplicities. We will return to discuss the 
yields of $\Omega,\overline\Omega$ at the end of this subsection.

In second last  column, the  superscript `s' 
means that $\lambda_{\rm s}$ is fixed by strangeness balance and,  superscript 
`$\gamma_{\rm q}$', in two last columns,  means that 
$\gamma_{\rm q}=\gamma_{\rm q}^{\rm c}=e^{m_\pi/2T_{\rm f}}$, 
is fixed to maximize the entropy content in the hadronic phase space.
The fits presented  are obtained with latest NA49 experimental results, \ie,
have updated  ${\rm h}^-/b$, newly published $\phi$ yield \bcite{App99}, and we predict
the ${\bar\Lambda}/{\rm \bar p}$ ratio.  $b$ is here the number of baryon
participants, and  $\rm h^-=\pi^-+K^-+\bar p$ is the yield of stable negative
hadrons which includes pions, kaons and antiprotons.
We see, comparing the two columns, that strangeness conservation 
(enforced in second last column) 
is consistent with the experimental data shown,  enforcing it 
does not change much the results for particle multiplicities. 

\begin{table}[tb]
\captt{resultpb2}{Pb--Pb 158\agev particle ratios}{
\small
WA97 (top) and NA49 (bottom)  Pb--Pb 158\agev collision hadron ratios
compared with phase space fits.
\index{particle ratios!Pb--Pb at 158\agev}}
\vspace{0.2cm}
\begin{center}
\begin{tabular}{lclll}
\hline\hline
\baselineskip 0.9cm
 Ratios & Ref. &  Exp. Data       & Pb$|^{\rm s,\gamma_{\rm q}}$ & Pb$|^{\rm \gamma_{\rm q}}$ \\
\hline
${\Xi}/{\Lambda}$ &  \protect\cite{Kra98} &0.099 $\pm$ 0.008                     & 0.096 & 0.095\\
${\overline{\Xi}}/{\bar\Lambda}$ &  \protect\cite{Kra98} &0.203 $\pm$ 0.024      & 0.197 & 0.199\\
${\bar\Lambda}/{\Lambda}$  &  \protect\cite{Kra98} &0.124 $\pm$ 0.013            & 0.123 & 0.122\\
${\overline{\Xi}}/{\Xi}$  &  \protect\cite{Kra98} &0.255 $\pm$ 0.025             & 0.251 & 0.255\\
\hline
${\rm K^+}/{\rm K^-}$         &  \protect\cite{Bor97}         &  1.80$\pm$ 0.10         & 1.746  & 1.771 \\
${\rm K}^-/\pi^-$         &       \protect\cite{Sik99}&  0.082$\pm$0.012        & 0.082  & 0.080\\
${\rm K}^0_{\rm s}/b$       &  \protect\cite{Jon96}   & 0.183 $\pm$ 0.027       & 0.192  & 0.195\\
${h^-}/b$                 &  \protect\cite{App99}     & 1.97 $\pm $ 0.1\ \      & 1.786  & 1.818 \\
$\phi/{\rm K}^-$   &  \protect\cite{Afa00}  & 0.145 $\pm$ 0.024\ \                 & 0.164  & 0.163 \\
${\bar\Lambda}/{\rm \bar p}$     &       $y=0$             &                       & 0.565  & 0.568 \\
${\rm \bar p}/\pi^-$             &       all $y$           &                       & 0.017  & 0.016 \\
\hline
 & $\chi^2$     &     & 1.6 & 1.15 \\
 &  $ N;p;r$     &    & 9;4;1& 9;5;1\\
\hline\hline
\end{tabular}
\end{center}
\end{table}
The six  parameters ($T, v_{\rm c}, \lambda_{\rm q}, 
\lambda_{\rm s}, \gamma_{\rm q}, \gamma_{\rm s}$)
describing the  particle abundances
are shown in the top section of \rt{fitqpbs}.
Since the results of the WA97 experiment  are not covering the full phase space, 
there is a reasonably precise value found  for  one velocity parameter,
 taken to be the spherical surface flow velocity $v_{\rm c}$ of the
fireball hadron source.

As in  S-induced reactions where $\lambda_{\rm s}=1$ \bcite{Let99},
now  in Pb-induced reactions, a value $\lambda_{\rm s}^{\rm Pb}\simeq 1.1$
characteristic for a source of freely movable  strange quarks with
balancing strangeness
in presence of strong Coulomb potential 
\cite[page 3568]{Let99d}, {\ie}, with $\tilde\lambda_{\rm s}=1$, is obtained. 
Since all chemical non equilibrium studies of the Pb--Pb 
system converge to the case of maximum entropy, 
we have  presented the results with fixed 
$\gamma_{\rm q}=\gamma_{\rm q}^{\rm c}=e^{m_\pi/2T_{\rm f}}$. 
The large values of $\gamma_{\rm q}>1$, seen in \rt{fitqpbs}, 
confirm the need to hadronize the excess entropy of the 
QGP possibly formed. This value is derived from both 
 the specific  negative hadron  ${\rm h}^-/b$ abundance  
and from the relative strange hadron yields.

\begin{table}[tb]
\captt{fitqpbs}{statistical model parameters}{
\small
Upper section: statistical model parameters
which best describe the experimental results for
Pb--Pb data seen in Fig.\,\ref{Tab:resultpb2}.
Bottom section: energy per entropy, antistrangeness, net strangeness
 of  the full hadron phase space characterized by these
statistical parameters. In column two, we fix $\lambda_{\rm s}$ by requirement of 
strangeness conservation, and in this and next column we fix 
$\gamma_{\rm q}=\gamma_{\rm q}^{\rm c}$.
Superscript $^*$ indicates values which are result of a constraint.
\index{chemical parameters!Pb--Pb system}
}
\vspace{0.2cm}\begin{center}
\begin{tabular}{lcc}
\hline\hline
                           & Pb$|_v^{\rm s,\gamma_{\rm q}}$ & Pb$|_v^{\rm \gamma_{\rm q}}$\\
\hline
$T$ [MeV]             &  151 $\pm$ 3      &  147.7 $\pm$ 5.6           \\
$v_{\rm c}$           & 0.55 $\pm$ 0.05   & 0.52 $\pm$ 0.29        \\
$\lambda_{\rm q}$     & 1.617 $\pm$ 0.028 & 1.624 $\pm$ 0.029       \\
$\lambda_{\rm s}$     & 1.10$^*$         & 1.094 $\pm$ 0.02         \\
$\gamma_{\rm q}$  & ${\gamma_{\rm q}^{\rm c}}^*=e^{m_\pi/2T_{\rm f}}$=1.6  &${\gamma_{\rm q}^{\rm c}}^*=e^{m_\pi/2T_{\rm f}}$=1.6\\
$\gamma_{\rm s}/\gamma_{\rm q}$   & 1.00 $\pm$ 0.06  & 1.00 $\pm$ 0.06         \\
\hline 
$E/b$[{\small GeV}]   &  4.0    &     4.1      \\
${s}/b$               & 0.70 $\pm$ 0.05  & 0.71 $\pm$ 0.05         \\
$E/S$[{\small MeV}]   & 163 $\pm$ 1    & 160 $\pm$ 1           \\
$({\bar s}-s)/b\ \ $  & 0$^*$            &  0.04 $\pm$ 0.05    \\ 
\hline\hline
\end{tabular}
\end{center}
\end{table}

The fits shown in \rt{fitqpbs} satisfy comfortably the constraint 
that $E/S>T$ discussed in \ress{SudHad}. 
One of the interesting  quantitative results of this analysis 
is shown in the bottom section of \rt{fitqpbs}: the
 yield of strangeness per baryon, $s/b\simeq 0.7$\,.
 We see, in lower portion of   \rt{fitqpbs}, that near strangeness 
balance  is obtained as result of the fit. 

The most rarely produced hadron is the triply strange 
$\Omega(\rm sss)$ and  $\overline\Omega(\rm \bar s\bar s\bar s)$ which are 
the heaviest stable hadrons, $M_\Omega=1672$\,MeV. The phase space for 
$\overline\Omega$ is ten times smaller than that for  $\overline\Xi$  at the 
conditions of chemical  freeze-out we have obtained, and
any non-statistical hadronization contribution  would be 
first visible in the\index{statistical hadronization!omega excess}
$\Omega$ and $\overline\Omega$ production pattern. 
For the parameters as  in \rt{fitqpbs},
the $\overline\Omega$ yields reported 
by the experiment WA97 are under predicted 
by nearly factor two. This yield  excess originates
at lowest $m_\bot$.
The `failure' of a statistical hadronization model to 
describe $\overline\Omega$ (and by 30\% $\Omega$) yields 
has several possible explanations. 

$\Omega$ and $\overline\Omega$ 
enhancement could be caused by strangeness 
pre-clustering in the deconfined phase \bcite{Raf82a}.
This would enhance all multistrange 
hadrons, but most prominently the phase space 
suppressed $\Omega$ and $\overline\Omega$ yields.  This mechanism 
would work only if pairing of strange quarks would be significant near to phase 
transition.
There is the possibility that distillation of strangeness followed by break up of
 strangelets which process 
could contribute to $\Omega$ and $\overline\Omega$ production.
The decay of disoriented chiral condensates has also been considered,
 \bcite{Kap01}.

In view of these pre and post-dictions of the 
$\Omega$ and $\overline\Omega$ anomalous yield,
one should abstain from introducing these particles into 
statistical hadronization model fits.
We note that the early statistical descriptions of 
$\Omega$ and $\overline\Omega$ yields have not been sensitive to the
problems we described \bcite{Bec98,Let97b}. In fact, as
long as the parameter $\gamma_{\rm q}$ is not considered, it is not possible
to describe the experimental data at the level of precision that
would allow recognition of the  $\Omega$ and $\overline\Omega$ excess yield
within statistical hadronization. For example, a chemical
equilibrium fit, which includes the  $\Omega$ and $\overline\Omega$ yield
has for 18 fitted data points with two parameters a 
$\chi^2/{\rm dof}=37.8/16$ \bcite{Bra99}. Such a fit is 
quite unlikely to contain all the physics even if its 
appearance  to untrained eye is suggesting a very good 
description of experimental data.

\subsection{Strangeness at RHIC}
\lssec{SRHIC}
In the likely event that the QGP formed at RHIC evolves
towards strangeness chemical 
equilibrium abundance, or possibly even exceeds it, we should 
expect a noticeable over occupancy  of strangeness as measured in terms 
of chemical equilibrium  final state hadron abundance. Because much of the
strangeness is in the baryonic degrees of freedom, the  kaon to pion ratio 
should appear suppressed,  compared to SPS results.
A more penetrating effect of the hadronization of strangeness rich 
QGP at RHIC is the formation of strange 
baryons and antibaryons. This high phase space occupancy is
one of the requirements for  the enhancement of multistrange (anti)baryon
production, which is an important hadronic signal of
QGP phenomena \bcite{Raf81b,Raf82c,Raf82a,Raf82b}. In particular, we
hope that hadrons produced in phase space with a small probability, 
such as $\Omega,\,\overline\Omega$, will be observed with a 
yield above these expectations, continuing the trend seen at SPS.

Many results from RHIC $\sqrt{s_{\rm NN}}=130$ GeV run are still
preliminary and the following quantitative 
discussion is probably not the final 
word in this matter. However, the results we find are very 
interesting, and in qualitative agreement with the sudden QGP break up
reaction picture. The data are mainly obtained
at the central rapidity region where, due to approximate  
longitudinal scaling, the effects of flow cancel and 
we can evaluate the full phase space yields
in order to obtain particle ratios. We do not explore trivial results
such as $\pi^+/\pi^-=1$, since the large hadron yield combined with the
flow of baryon isospin asymmetry towards the fragmentation rapidity region 
assures us that this result will occur to a great precision.
We also do not use the results for
$\rm K^*, \bar K^*$ since these yields depend on the degree of rescattering of
resonance decay products \bcite{Tor01,Raf01}. 
The data we use has been reported in conference reports
 of the STAR collaboration of Summer 2001, which as available 
are combined with data of PHENIX, BRAHMS, PHOBOS, for more
discussion of the data origin, see \bcite{Bra01}.
We assume, in our fit in \rt{RHIChad}, that the multistrange
weak interaction cascading $\Xi\to \Lambda$, in the STAR
result we consider,  is  cut by  vertex discrimination
and thus we use these yields without weak interaction corrections.

\begin{table}[t]
\captt{RHIChad}{RHIC 130 analysis}{ 
\small
Fits of central rapidity hadron ratios for RHIC  $\sqrt{s_{\rm NN}}=130$ GeV run.
Top section: experimental results, followed by chemical parameters, 
physical property of the phase space, and the fit error. Columns: data, full non-equilibrium
fit, nonequilibrium fit constrained by strangeness conservation and supersaturation 
of pion phase space,  and in the last column, equilibrium fit constrained by 
strangeness conservation, upper index $^*$ indicates quantities fixed by these 
considerations.
\index{particle ratios!Pb--Pb at $\sqrt{s_{\rm NN}}=130$ GeV }
\index{chemical parameters!Pb--Pb at $\sqrt{s_{\rm NN}}=130$ GeV}}
\vspace*{0.2cm}
\begin{center}
\begin{tabular}{lcccc}
\hline\hline
                                        & Data  & Fit& Fit         & Fit$^{\rm eq}$   \\
                                        &       &    & $\rm s-\bar s=0$&  $\rm s-\bar s=0$\\
\hline
$ \rm{\bar p}/p$                            &0.64\ $\pm$0.07\ & 0.637 & 0.640 &  0.587 \\
$\rm{\bar p}/h^-$                          &                & 0.068 & 0.068 &  0.075 \\
${\overline\Lambda}/{\Lambda}$          &0.77\ $\pm$0.07\ & 0.719 & 0.718 &  0.679  \\
$\rm{\Lambda}/{h^-}$                       & 0.059$\pm$0.001 & 0.059 & 0.059 &  0.059  \\
$\rm{\overline\Lambda}/{h^-}$              & 0.042$\pm$0.001 & 0.042 & 0.042 &  0.040  \\
${\overline{\Xi}}/{\Xi}$                &0.83\ $\pm$0.08\ & 0.817 & 0.813 &  0.790  \\
${\Xi^-}/{\Lambda}$                     & 0.195$\pm$0.015 & 0.176 & 0.176 &  0.130  \\
${\overline{\Xi^-}}/{\overline\Lambda}$ & 0.210$\pm$0.015 & 0.200 & 0.200 &  0.152  \\
$\rm{K^-}/{K^+}$                           &0.88\ $\pm$0.05\ & 0.896 & 0.900 &  0.891  \\
$\rm{K^-}/{\pi^-}$                         & 0.149$\pm$0.020 & 0.152 & 0.152 &  0.145  \\
$\rm{K_S}/{h^-}$                           & 0.130$\pm$0.001 & 0.130 & 0.130 &  0.124  \\
${\Omega}/{\Xi^-}$                      &                & 0.222 & 0.223 &  0.208 \\
${\overline{\Omega}}/{\overline{\Xi^-}}$&                & 0.257 & 0.256 &  0.247 \\
${\overline{\Omega}}/{\Omega}$          &                & 0.943 & 0.934 &  0.935    \\
\hline
$T$                                       &           &158$\pm$ 1   & 158$\pm$ 1     &  183$\pm$ 1      \\
$\gamma_{\rm q}$                            &      &1.55$\pm$0.01 & 1.58$\pm$0.08  &  1$^*$    \\
$\lambda_{\rm q}$                           &      &1.082$\pm$0.010& 1.081$\pm$0.006& 1.097$\pm$0.006      \\
$\gamma_{\rm s}$                            &       & 2.09$\pm$0.03 &  2.1$\pm$0.1   &  1$^*$    \\
$\lambda_{\rm s}$                           &     &$\!$1.0097$\pm$0.015$\!$&  1.0114$^*$     & 1.011$^*$ \\
\hline
$E/b$[{\small GeV}]\ \                        &   &  24.6& 24.7  &  21    \\
$s/b$                                         &   &  6.1 &  6.2  &  4.2  \\
$S/b$                                         &   & 151  &  152  &  131  \\
$E/S$[{\small MeV}]\ \                        &   & 163  &  163  &  159  \\
\hline
$\chi^2/$dof                                 &   & 2.95/($10\!-\!5$)  &2.96$\!$/$\!$($10\!-\!4$)  & 73/($10\!-\!2$)  \\
\hline\hline
\end{tabular}
\end{center}
\end{table}

We first look at the last column in  \rt{RHIChad},
the chemical equilibrium fit. Its large  $\chi^2$
originates in the inability to account for multistrange 
$\overline\Xi,\,\Xi$. Similar results are 
presented in Ref.\bcite{Bra01}, in an equilibrium fit 
which  does not include multistrange hadrons.
The equilibrium fit yields $E/S=159\,\mbox{MeV}<T=183$ MeV
contradicting the conditions we discussed in depth in \ress{SudHad}.
On the other hand, the chemical nonequilibrium fits come out
to be in perfect agreement with data, and are consistent with the QGP hadronization
picture since  $E/S=163>T=158$ MeV and $\gamma_{\rm s},\gamma_{\rm q}>1$\,.
The value of the hadronization temperature $T=158$ MeV is 
 below the  central expected equilibrium phase transition
temperature, this  hadronization  temperatures at RHIC is  
consistent with sudden breakup of a supercooled QGP fireball.
The inclusion of the  
yields of multistrange antibaryons in the RHIC data analysis, 
along with  chemical non-equilibrium, allows to
discriminate the different reaction scenarios. 

We look next at the strangeness content, $s/b=6$, in  \rt{RHIChad}: the full
QGP phase space would have yielded  8.6 strange quark pairs per baryon
at $\lambda_{\rm q}=1.085$,  as we will show below in \rf{PLSBLAMQ}. Thus we
conclude that $\gamma_{\rm s}^{\rm QGP}=6/8.6=0.7$.  With this value, and using the
fitted value  $\gamma_{\rm s}^{\rm HG}=2.1$, we compute 
$\gamma_{\rm s}^{\rm HG}/\gamma_{\rm s}^{\rm QGP}=2.1/0.7=3$.   The fact that the 
strangeness phase space in QGP is not fully saturated is,
on a second careful look, in qualitative agreement
with  kinetic theory predictions, adjusting for the observed RHIC-130 run
conditions. 

The  value of the thermal energy content, $E/b=25$ GeV, seen in \rt{RHIChad} is also in very good 
agreement with expectations once we allow for the kinetic energy content,
associated with longitudinal and transverse motion. The energy of each particle 
is `boosted' with the factor $\gamma_\bot^v\cosh y_\parallel$. 
For $v_\bot=c/\sqrt{3}$, we have $\gamma_\bot^v=1.22$.
The longitudinal flow range is about $\pm 2.3$ rapidity units, 
according to PHOBOS results.
To obtain the the energy increase due to longitudinal flow, we have to multiply 
by the average,  
$\int dy_\parallel \cosh y_\parallel/y_\parallel\to\sinh(2.3)/2.3=2.15$, 
for a total average increase in energy by 
factor 2.62, which takes the full energy content to 
$E^v/b\simeq 65$ GeV as expected.

We now consider what  experimental hadron yield results imply about 
total strangeness yield in the RHIC fireball. 
First,  we sum up the yield of strange quarks contained in hyperons. 
We have in singly strange hyperons 1.5 times
the yield observed in $\Lambda$,
since $\Sigma^{\pm}$ remain unobserved. Also, accounting for the doubly strange
$\Xi^-$ which are half of the all $\Xi$, and contain two strange quarks, we have:
\[
{\langle s\rangle_{\rm Y}\over\rm  h^-}=1.5\cdot 0.059+2\cdot 2\cdot  0.195\cdot 0.059=0.133\,.
\]
Allowing for the unobserved $\Omega$ at the theoretical rate,
this number increases to $\langle s\rangle_{\rm Y}/\rm h^- =0.14$. Repeating the same 
argument for antihyperons the result is $0.10$. $\rm s$ and $\rm \bar s$ content in kaons is 
four times that in $\rm K_S$ and thus the total strangeness yield is
\[
{\langle\rm  s + \bar s\rangle\over\rm  h^-}=0.76\,,
\]
with 32\% of this yield contained in hyperons and antihyperons. Up to this
point, the analysis is  based on direct measurements and
established particle yields. 

We now estimate the increase in the `strangeness suppression' factor $W_{\rm s}$, \req{wro1}. 
Correcting for presence of $\rm K^-$ among  negatively charged hadrons, and 
assuming that all three pions are equally abundant, we find: 
\[
{\rm \langle s + \bar s\rangle\over \pi^-+\pi^+ +\pi^0}\simeq 0.30\,.
\]
The total number of pions produced comprises pions arising from resonance
decays and from projectile and target fragments. Thus, as little as half
of the pions are originating in the newly made $\rm q\bar q$ pairs.   
In the RHIC $\sqrt{s_{\rm NN}}=130$ GeV run, $W_{\rm s}\simeq 0.6$.
\index{strangeness!yield at RHIC}
The increase compared to SPS 
is largely due to strangeness content in hyperons.
Considering that $\gamma_{\rm s}^{\rm QGP}\simeq 0.7$ at $\sqrt{s_{\rm NN}}=130$ GeV,
there is still space for a further strangeness yield rise at highest RHIC energy,
and we hope and  expect $W_{\rm s}\to 1$ when initial temperatures rises to well above
strange quark mass for sufficient length of time.

\subsection{Strangeness as signature of deconfinement}
\lssec{SsigQGP}\lssec{Syield}
The rate of strangeness production in QGP
is sensitive to the temperature achieved
at the time gluons reach chemical and thermal  equilibrium. 
There is considerable uncertainty how short the time required to
relax strangeness flavor in the thermal gluon medium is. Consideration 
of a small strangeness mass
found in lattice studies of strange hadrons implies rapid strangeness 
chemical equilibration. There is
also the probable acceleration of equilibration due to the next to leading order effects
($K$-factor). In view of this, we now discuss a benchmark yield 
of strangeness, assuming that the equilibration process 
leads to near chemical equilibrium conditions for hadronizing QGP.
Specifically, the light quark abundance in the \QGP phase is considered at 
the  equilibrium yield, while the strange quark yield is characterized by
the QGP-phase space occupancy before hadronization, $\gamma_{\rm s}^{\mathrm{QGP}}$.

We consider the ratio of equilibrium strangeness density, arising in the
Boltzmann gas limit,  to the baryon density 
in a QGP fireball  yields:

\beql{sdivb}
\frac{\rho_{\rm s}}{\rho_{\mathrm b}}={s\over b}=
\frac{\gamma_{\rm s}^{\mathrm{QGP}} {3\over \pi^2} T^3 
(m_{\rm s}/T)^2K_2(m_{\rm s}/T)}
  {\frac23\left(\mu_{\rm q} T^2+{\mu_{\rm q}^3/ \pi^2}\right)}\,.
\eeq
To first approximation,
perturbative thermal QCD corrections cancel in the ratio.
For $m_{\rm s}=200$ MeV and $T=150$ MeV, we have:
\beql{sdivb1}
{s\over b}\simeq  
  \gamma_{\rm s}^{\mathrm{QGP}}\frac{0.7}
          {\ln \lambda_{\rm q} +{{(\ln \lambda_{\rm q})^3}/{\pi^2}}}\,.
\eeq

The relative yield $s/b$ is 
in the approximation considered, nearly temperature independent, which 
allows to  gain considerable understanding of strangeness production. This 
ratio is mainly dependent on the value of $\lambda_{\rm q}$. This
light quark fugacity pertinent to the final state hadrons 
is usually quite  well determined and does not 
vary depending on the strategy of data analysis. In the 
sudden hadronization model, the value in QGP is imprinted on the
yields of final state hadrons.

\myfig{0.3}{12}{PLSBLAMQB}{0}{-4.5}{-1.6}{tb}
\capt{PLSBLAMQ}{S/b yield}{
\small
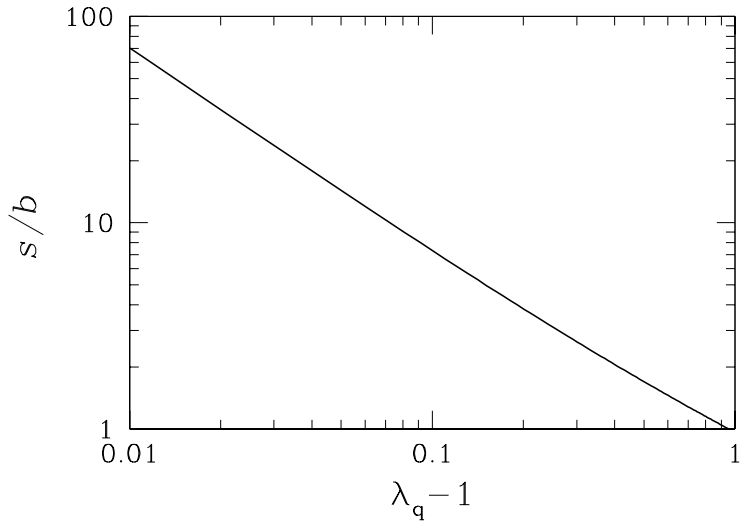
Strangeness yield per baryon as function of $\lambda_{\rm q}$ 
in equilibrated quark-gluon plasma.
}
\end{figure}

We show in \rf{PLSBLAMQ}, as function of 
$\lambda_{\rm q}-1$ (variable chosen to enlarge the interesting region
$\lambda_{\rm q}\to 1$), the expected relative yield per baryon
originating in the QGP, defined in \req{sdivb1} with 
$ \gamma_{\rm s}^{\mathrm{QGP}}=1$. 
At  top SPS energy, we see that 
the equilibrium yield is at 1.5 strange pairs per participating baryon 
(for $\lambda_{\rm q}\simeq 1.5$--$1.6$).
In p--p collisions at the corresponding energy, the yield is below 0.3 strange
pair per participant \bcite{Wro85}.  We note that the  p--p reaction
yield is suppressed by factor 2 compared to the canonical yield resulting
when one attempts to use statistical canonical description to size up
the phase space.
The actual experimental yield, is 2.5 times the
yield in p--p reactions, see \rt{fitqpbs}, this is however  only half as large
as  in an equilibrated QGP,  see  \rf{PLSBLAMQ}. There is need to have 
$\gamma_{\rm s}^{\mathrm{QGP}}\simeq 0.5$--0.7, in both  p--p reaction and the SPS 
top energy Pb--Pb reactions. The explanation of this is that the QGP
system did not get to be hot enough for long enough time
to fill the small p--p and very large Pb--Pb phase spaces.

At the RHIC 130 GeV run, the value $\lambda_{\rm q}=1.09$
allows to understand many particle yields at central rapidity. 
We see, in \rf{PLSBLAMQ}, that specific strangeness yield in a
QGP fireball at equilibrium is 
expected to be an order of magnitude greater than 
currently observed at SPS top energy. However, 
comparing to general hadron multiplicity, only a factor 1.5--2 
further strangeness enhancement can at most be expected at RHIC,
the remarkable feature of the RHIC situation is that this enhancement is 
found in the (multistrange) baryon abundance. Specially, given the large 
strangeness per baryon ratio, \rf{PLSBLAMQ},  baryons and antibaryons produced at
RHIC are mostly  strange \bcite{Raf99a}. We are not aware of any 
reaction model other than QGP formation and hadronization which 
could produce this type of anomaly.

While the specific strangeness yield $s/b$ is a clear indicator for
the extreme conditions reached in heavy ion collisions, perhaps an equally
interesting observable is the occupancy of the hadron 
strangeness phase space, $\gamma_{\rm s}^{\mathrm{HG}} $. The interesting
result to expect is an enhancement due to the need to hadronize,
into a strangeness poor phase, the QGP strangeness excess. To see this,
we compare the phase space of strangeness in QGP with that of resulting HG. 
The absolute yields must be the same in both phases. This hadronization 
condition allows to relate the two phase space occupancies in 
HG and QGP, by equating the strangeness content in both phases. 
Canceling the common normalization factor $T^3/(2\pi^2)$, we obtain: 
\beql{stranHAD}
\gamma_{\rm s}^{\mathrm{QGP}} V^{\mathrm{QGP}}g_{\rm s} 
\left({m_{\rm s}\over T^{\mathrm{QGP}}}\right)^2
K_2\left({m_{\rm s}\over T^{\mathrm{QGP}}}\right)\simeq
\gamma_{\rm s}^{\mathrm{HG}}  V^{\mathrm{HG}} \left[{ \gamma_{\rm q}\lambda_{\rm q}\over \lambda_{\rm s}}F_{\rm K}
          \!+\!{\gamma_{\rm q}^2\over \lambda_{\rm q}^{2}\lambda_{\rm s}}F_{\rm Y}\right],
\eeq
where $F_{\rm K}$ and $F_{\rm Y}$ are the phase spaces of kaons and hyperons, respectively.
We have, without loss of generality, 
  followed the $\rm\bar s$ carrying hadrons in the
HG phase space, and we have,
in first approximation,  omitted the contribution of multistrange
antibaryons. We now use the condition that strangeness is
conserved to eliminate $\lambda_{\rm s}$,   in \req{stranHAD}, and obtain,
\beql{stranHAD1}
{\gamma_{\rm s}^{\mathrm{HG}}\over \gamma_{\rm s}^{\mathrm{QGP}}} 
{V^{\mathrm{HG}}\over V^{\mathrm{QGP}}}
  =
{{g_{\rm s} W(m_{\rm s}/T^{\mathrm{QGP}})}\over{ \sqrt{
(\gamma_{\rm q} F_{\rm K}+ \gamma_{\rm q}^2 \lambda_{\rm q}^{-3}F_{\rm Y})
(\gamma_{\rm q} F_{\rm K}+ \gamma_{\rm q}^2 \lambda_{\rm q}^{3}F_{\rm Y})}}}\,.
\eeq

\myfig{0.3}{12}{PLGAHGQGPB}{0}{-4.5}{-1.6}{tb}
\capt{PLGAHGQGP}{gammaS ratio HG to QGPas function of lambdaQ}{
\small
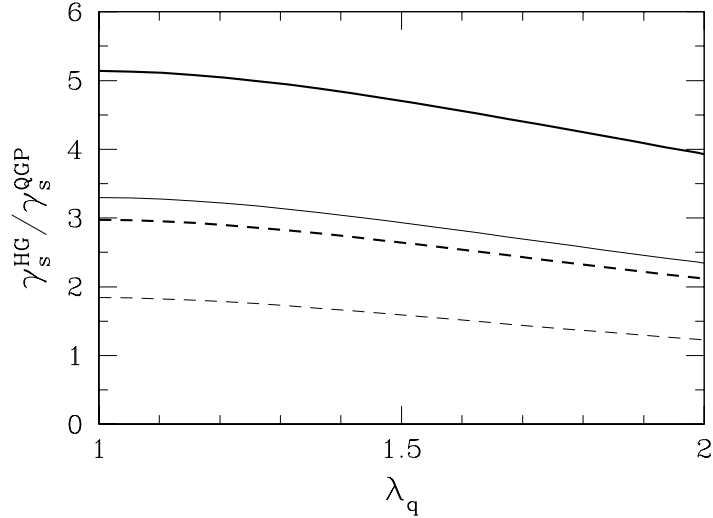
Strangeness occupancy $\gamma_{\rm s}$  ratio HG/QGP in sudden hadronization
as function of  $\lambda_{\rm q}$. Solid lines $\gamma_{\rm q}=1$, 
 and short dashed $\gamma_{\rm q}=1.6$.
Thin lines for $T=170$ and thick lines $T=150$ MeV, common to both 
phases. 
}
\end{figure}

In sudden hadronization,  
$V^{\mathrm{HG}}/ V^{\mathrm{QGP}}\simeq 1$, 
the growth of volume is negligible,  
$T^{\mathrm{QGP}}\simeq T^{\mathrm{HG}}$, the temperature 
is maintained across the hadronization front, and the chemical occupancy 
factors in both states of matter accommodate the different magnitude of the particle
phase space. In this case, the QGP  strangeness when `squeezed' into the
smaller HG phase space results in 
${\gamma_{\rm s}^{\mathrm{HG}}/\gamma_{\rm s}^{\mathrm{QGP}}}> 1$\,.
We show, in \rf{PLGAHGQGP}, the enhancement of phase space occupancy 
expected in sudden hadronization of the QGP. The temperature range 
$T=150$ MeV (thick lines) and $T=170$ MeV (thin lines) spans the 
range considered today at SPS and RHIC. The value of $\gamma_{\rm q}$ is in
range of the chemical equilibrium in HG,  $\gamma_{\rm q}=1$
(solid lines),  to the expected excess in sudden hadronization,
$\gamma_{\rm q}=1.6$ (short dashed lines). 

We note that, for the top SPS energy range where 
$\lambda_{\rm q}=1.5$--1.6, sudden hadronization analysis of data implies 
$T\simeq 150 \mbox{\,MeV},\,\gamma_{\rm q}\simeq 1.6$, and the value of $\gamma_{\rm s}$ increases
across hadronization by factor 2.7\,.  Since the yield of strangeness seen at SPS
implies $\gamma_{\rm s}^{\mathrm{QGP}}\simeq 0.6$, this in turn implies 
$\gamma_{\rm s}^{\mathrm{HG}}\simeq 1.6\simeq \gamma_{\rm q}^{\mathrm{HG}}$, as
is indeed found in hadron production analysis in the sudden hadronization
picture, \rt{fitqpbs}. Because accidentally 
$\gamma_{\rm s}^{\mathrm{HG}}/ \gamma_{\rm q}^{\mathrm{HG}}\simeq 1$,
one can also model the hadronization at SPS energy in terms of an
equilibrium hadronization model. The pion enhancement associated with 
the high entropy phase can be accommodated by use of two temperatures, one
for the determination of absolute particle yields, and another for 
determination of the spectral shape. Such an approach has
 similar number of parameters, and comparable predictive power.

However, the SPS condition, 
$\gamma_{\rm s}^{\mathrm{HG}}/ \gamma_{\rm q}^{\mathrm{HG}}\simeq 1$,
is  not present at the  RHIC energy range, where the
hadron phase space occupancy for strangeness is significantly larger than
for light quarks, see  \rt{RHIChad}. QGP hadronization dynamics should 
emerge clearly from the study of (multi)strange hadron yields at RHIC.
Using the yields of (multi) strange baryons and antibaryons
it is possible to discriminate against trivial 
equilibrium hadronic gas models.  Moreover,
at RHIC we  find a very strong strange baryon and antibaryon yield 
which  is not accessible in kinetic parton models.

\section{FINAL REMARKS}
The deconfined thermal phase manifests itself through  its gluon content, 
which generates in thermal collision processes a clear strangeness 
fingerprint of QGP.
The SPS  strangeness results decisively show  interesting new 
physics, with a significant excess of strangeness and
strange antibaryons, and spectral symmetry between baryons 
and antibaryons.  We see, at SPS and at RHIC, 
considerable convergence of the hadron production 
around properties of suddenly hadronizing entropy and
strangeness rich  QGP.
We see hadronization into pions, at  
$\gamma_{\rm q}\to \gamma_{\rm q}^{\rm c}=e^{m_\pi/2T_{\rm f}}\simeq 1.6$,
which is an  effective way to convert excess of entropy 
in the plasma into hadrons. Because of large value of
$\gamma_{\rm q}$ the
strange particle signature of QGP hadronization 
become more extreme and clear at RHIC, as the strangeness
 excess is covered by the general hadron yield excess at SPS.
The systematic behavior as function of collision energy, and 
other collision system parameters, will in future provide 
a cross check of our reaction picture.

We  have not  advocated that 
the understanding of QGP can or should 
be based on a comparison  of the  A--A collision system
with the elementary systems such as p--p or even $\rm e^+$--$\rm e^-$. 
We simply do not know at this time if elementary 
interaction system is in any fundamental way different. It could well 
be that once there is enough energy deposition,
 the relativistic Maxwell demon which leads to the formation of the
statistical high entropy QCD state is operating in both the small 
and the big system, and that the difference in the observables
arises due to a change in the internal excitation
(temperature), and size (and
thus lifespan). For example, there would be no outward (transverse)
flow of matter and little strangeness thermal production 
expected in p--p system, compared to  A--A collision system.
The smaller phase space of the  p--p system is easier to equilibrate 
than the  A--A collision system, leading to further confusion.

It is important to remember that, not only at RHIC, and 
in near future at LHC, 
QGP can be studied. It is very probable that the onset of 
deconfinement occurs at quite modest energies, perhaps in 
collisions of 20--40$A$ GeV Pb projectiles with a 
laboratory target.  An alternate method of QGP
exploration is  the study of the energy domain near the 
transition. Formation of QGP phase is an endothermic process 
with onset of entropy and strangeness production, and in
experiments near to the condition for phase
transformation, one should be able to recognize these properties 
quite clearly: for example, by the onset as function of energy
and of reaction volume of multistrange antibaryon
production. The measurement of yields as function
of energy, and volume (excitation functions) should
provide information about the nature of the transition
to this new phase of matter.

In closing, we have analyzed available RHIC and 
high energy SPS results and have demonstrated that these can be
easily understood as result of sudden hadronization of a rapidly 
expanding QGP fireball. We have discussed characteristic nonequilibrium aspects
of strangeness yields associated with the QGP fireball explosion.
Our analysis shows in detail how and where the hadronic equilibrium
models fail to describe the experimental data. 

\end{document}